\documentclass[journal]{IEEEtran}%
\normalsize

\ifCLASSINFOpdf

\else

\fi
\usepackage{soul}
\usepackage{array}
\usepackage{color}
\usepackage{amsfonts}
\usepackage{amssymb}
\usepackage{amsmath}
\usepackage{epstopdf}
\usepackage[dvips]{graphicx}
\usepackage{cite}
\usepackage{balance}
\usepackage{caption}
\usepackage{subcaption}
\usepackage{textcomp}
\usepackage{gensymb}
\usepackage{float}
\usepackage{tabularx}
\usepackage{multirow}
\usepackage{amsthm}
\newcommand\underrel[3][]{\mathrel{\mathop{#3}\limits_{%
      \ifx c#1\relax\mathclap{#2}\else#2\fi}}}

\begin{document}
\title{Effect of Impulsive Noise on Uplink NOMA Systems 
}
\author{Bassant Selim, \IEEEmembership{Member, IEEE}, Md Sahabul Alam, \IEEEmembership{Student Member, IEEE},\\ Georges Kaddoum, \IEEEmembership{Member, IEEE} and Basile L. Agba, \IEEEmembership{Senior Member, IEEE} 
\thanks{B. Selim and G. Kaddoum are with the Electrical Engineering Department, ETS, University of Quebec, Montreal, QC H3C 1K3, Canada (emails: bassant.selim.1@ens.etsmtl.ca; \,\,georges.kaddoum@etsmtl.ca).}
\thanks{ M. S. Alam is with the Systems and Computer Engineering Department, Carleton University, Ottawa, ON K1S 5B6, Canada (email: sahabulalam@sce.carleton.ca) }
\thanks{B. L. Agba is with the Hydro-Qubec Research Institute (IREQ), Varennes, QC J3X 1S1, Canada (email: agba.basilel@ireq.ca).}
}
\maketitle	
\vspace{-5em}
\begin{abstract}
Non-orthogonal multiple access (NOMA) was recently proposed as a viable technology that can potentially provide the spectral efficiency, low latency, and massive connectivity requirements of future radio networks. In this context, numerous ultra-high reliability technologies such as the industrial Internet of things, smart grids, and smart homes present environments which are characterized by the presence of impulsive electromagnetic interference, known as impulsive noise. Under such conditions, the power domain multiplexing in NOMA is expected to render the system particularly sensitive to this additional impulsive noise. Therefore, in this article, we quantify the effects of impulsive noise on the outage performance of uplink NOMA systems. Extensive Monte-Carlo simulations as well as offered analytical results demonstrate the vulnerability of the involved NOMA users to this type of noise. This highlights the need for effective modeling of the impulsive noise as well as the design of mitigation techniques that are suitable for the particular demands and challenges of NOMA.
\end{abstract}
\section{Introduction}\label{S:Intro}
The concept of connecting every single device or "thing" to the Internet, namely the Internet of things (IoT), is continuously evolving encompassing more and more applications with a promise to revolutionize every aspect of our lives. Meanwhile, the ever-increasing applications of the aforementioned IoT technologies together with their stipulations present substantial challenges to the research community in order to address the distinct connectivity, reliability, and latency requirements of the massive number of connected devices. For instance, it is evident that conventional orthogonal multiple access schemes are relatively limited and incapable of supporting these highly demanding wireless systems and services. On this basis, non-orthogonal multiple access (NOMA) was introduced as a promising approach that is capable of overcoming the aforementioned challenges and therefore, rendering it a propitious candidate for future radio systems \cite{8010756}.

It was shown that NOMA can potentially provide substantially increased spectral efficiency, higher cell-edge throughput, and less stringent channel feedback, as only the received signal strength is required, see  \cite{7676258} and the references therein. Although there has been a growing literature on NOMA, particularly in the context of the IoT, most works consider a downlink scenario \cite{7973146}. In contrast to the downlink case, in uplink transmission, the outage probability (OP) analysis is a challenging exercise since the signal to noise ratio (SNR) comprises the channels of the different users, which are sorted and thus dependant. In this context, in \cite{7390209}, the authors formulated the OP and achievable sum data rate expressions of uplink NOMA transmission and derived the corresponding closed forms for the case of $2$ users. Similarly, in \cite{8309422}, the authors proposed an advanced multi-user decoding strategy based on the statistical channel state information (CSI) and derived the corresponding closed form OP for two users. Considering the general case of $M$ users, Liu \textit{et al.} \cite{8094298} and Wang \textit{et al.} \cite{8246494} resorted to approximations in order to derive the OP of uplink NOMA systems. Although instructive, all these works were built upon the classical assumption of additive white Gaussian noise (AWGN).
In fact, apart from some sporadic results \cite {8727449,7936664,8327593}, considering downlink scenarios of cooperative NOMA in power line communication systems, to the best of the authors' knowledge, the available literature on NOMA has exclusively considered this simplistic and sometimes unrealistic noise model. Indeed, several studies show sufficient evidences that impulse man-made noise is encountered in various metropolitan, high voltage, manufacturing plants, and indoor environments \cite{233212}. It was shown in \cite{1093943} that communication systems designed under the AWGN assumption typically experience severe performance degradation when subjected to impulsive noise. In addition, the power domain multiplexing in NOMA is expected to render such systems particularly sensitive to this additional impulsive noise \cite{8600314}. Thus a study of NOMA systems, which are not only disturbed by fading, but also by impulsive (non-Gaussian) noise, is of paramount importance in order to draw the practical limits of such systems and to provide pragmatic information for the system designer.

Therefore, this work seeks to fill this research gap and provide a practical perspective into the performance of power domain multiplexing NOMA-based systems under the effects of impulsive noise. To this end, this paper analyses the performance of uplink NOMA-based IoT systems, where the users, i.e. the IoT sensor units (SUs), employ NOMA to communicate with their respective access point (AP), under the effect of impulsive noise. We derive the exact outage probability and the asymptotic diversity order of an $M$ SUs uplink NOMA system subject to impulsive noise. It is shown that the power domain multiplexing in NOMA renders the users more sensitive to such noise than their orthogonal multiple access (OMA) counterparts.
\section{System Model}
We consider an uplink NOMA network consisting of an AP and $M$ SUs, with transceivers equipped with a single antenna. Let $h_i$ represent the instantaneous Rayleigh fading gain between the $i^{\rm th}$ SU and the AP.  Since the basic idea behind power domain NOMA is to superimpose the signals in the power domain and assuming that the AP utilizes the instantaneous CSI to determine the decoding  order\cite{7273963}, the received signal is given by
\begin{equation}
r=\sum\limits\limits_{i=1}^{M}\sqrt{{\rm{a}}_i}h_is_i+n,
\end{equation}
\noindent where the symbol index is dropped for notational convenience, the subscript $i$ refers to the $i^{\rm th}$ sorted SU while $s_i$ and ${\rm{a}}_i$ denote the information symbol and power of the $i^{\rm th}$ sorted SU, respectively. Without loss of generality, we assume that $\mathbb{E}\{|s_i|^2\}=1$. Moreover, $n=n_w+b n_I $ denotes the  total noise which is assumed to follow the widely used Bernoulli-Gaussian process \cite{ghosh1996analysis}, where $n_w$ is the background noise and $n_I$, the impulsive noise, both modelled as circularly symmetric complex Gaussian random variables with zero mean and variances $\sigma_w^2$ and $\sigma_I^2$, respectively. The parameter $\Gamma=\sigma_I^2/\sigma_w^2$ quantifies the impulsive to Gaussian noise power ratio. Finally, $b$ is the Bernoulli process, i.e., an independent and identically distributed (i.i.d.) sequence of zeros and ones that takes a value of $1$ with a probability $\textup{Pr}(b=1)=p$ and a value of $0$ with a probability $\textup{Pr}(b=0)=1-p$. Therefore, the probability distribution function (PDF) of the combined noise $n$ is given by \cite{ghosh1996analysis}
\begin{equation}
\begin{split}
f(n)=\frac{\left(1-p\right)}{{\pi \sigma_w^2}}e^{\frac{-|n|^2}{\sigma_w^2}}+\frac{p}{{\pi \left(\sigma_w^2+\sigma_I^2\right)}}e^{\frac{-|n|^2}{\sigma_w^2+\sigma_I^2}},
\end{split}
\end{equation}
\noindent where $\left|\cdot\right|$ denotes the absolute value operation. Let the channel gains of the SUs be sorted in ascending order as $|h_1|^2\leq|h_2|^2\leq...|h_j|^2\leq...|h_i|^2\leq...\leq|h_M|^2$, the AP will start by decoding the powerful $M^{\rm{th}}$ SU's signal and perform successive interference cancellation (SIC) to cancel the resulted interference, then it will proceed to decoding the $\left(M-1\right)^{\rm th}$ SU's signal and so on. Therefore, when decoding SU $j$'s message, the signals intended for all SUs $i$, where $i>j$, are canceled whereas the signals of the SUs with $i<j$ are treated as noise. Hence, assuming perfect cancellation and perfect CSI, the instantaneous SNR per symbol of the $j^{\rm th}$ sorted SU's message is given by
\begin{equation}
\label{eq:gamma_ideal}
\gamma_{j}=\left(1-p\right)\frac{a_j |h_j|^{2}}{\sum\limits\limits_{q=1}^{j-1}a_q|h_q|^{2}+{\rho_w}^{-1}}+p\frac{a_j |h_j|^{2}}{\sum\limits\limits_{q=1}^{j-1}a_q|h_q|^{2}+{\rho_I}^{-1}},
\end{equation}
\noindent where $|h_j|^{2}$ is the $j^{\rm th}$ sorted SU's instantaneous channel gain, $\rho_w=\frac{1}{\sigma_w^2}$ and $\rho_I=\frac{1}{\sigma_w^2+\sigma_I^2}=\frac{\rho_w}{\Gamma+1}$.
\section {Outage Probability Analysis}
 The OP can be defined as the probability that the symbol error rate is greater than a certain quality of service requirement and it can be computed as the probability that the SNR falls bellow a corresponding threshold. Assuming $j<M$, an outage event occurs when the AP cannot detect $\rm{SU}_j$'s signal or the signal of any $\rm{SU}_i$ in the SIC (i.e., $j<i\leq M$), which is formulated as \cite{7390209}
 \small
\begin {equation}
\begin{split}
\label{eq:POUT}
P_{(out,j)}&=\left(1-p\right)\left(1-\textup{Pr}\left\{E^c_{M|{w}}\bigcap \cdot \cdot \cdot\bigcap E^c_{j|{w}}\right\}\right)\\&\,\,\, +p\left(1-\textup{Pr}\left\{E^c_{M|{I}}\bigcap \cdot \cdot \cdot\bigcap E^c_{j|{I}}\right\}\right)\\
&=1-\left(1-p\right)\prod_{i=j}^{M}\textup{Pr}\left(E^c_{i|w}\right)-p\prod_{i=j}^{M}\textup{Pr}\left(E^c_{i|I}\right),
\end{split}
\end {equation}
\normalsize
\noindent where the subscript $i|s$, $s\in\{w,I\}$, refers to the $i^{\rm th}$ SU given the state of the noise, i.e. impulsive or background. Here, $E_{i|s}=\{R_{i|s}<{{R^t_{i}}} \}$ is the event that the $i^{\rm th}$ SU's message cannot be detected when the noise state is $s$ where $E_{i|s}$ and $E_{j|s}$ are mutually independent for $i\neq j$. Moreover, $R_{i|s}$ denotes the $i^{\rm th}$ SU's rate, ${{R^t_i}}$ is the targeted data rate of the $j^{\rm th}$ SU and $E_{i|s} ^c$ the complementary set of $E_{i|s}$ which is expressed as
\begin{equation}
\label{eq:cond_step1}
E^c_{i|s} =\left\{\frac{a_i |h_i|^{2}}{\sum\limits\limits_{q=1}^{i-1}a_q|h_q|^{2}+\frac{1}{\rho_s}}>\phi_{i}\right\},
\end{equation}
\noindent where $\phi_i=2^{{R^t_i}}-1$. Since the channels $|h_i|^{2}$ are ordered and thus, dependant, deriving $\textup{Pr}\left(E^c_{s|I}\right)$ involves integrating the cumulative distribution function of $|h_1|^{2}$ over the joint distribution $f_{|h_2|^{2},...,|h_i|^{2}}\left(|h_2|^{2},...,|h_i|^{2}\right)$  for $|h_1|^2\leq|h_2|^2\leq...|h_j|^2\leq...|h_i|^2\leq...\leq|h_M|^2$, which is intractable. Meanwhile, it has been established that for  $M$ i.i.d. exponentially distributed random variables, the distribution of the corresponding $i^{\rm th}$ ordered statistics coincides with the distribution of \cite{nevzorov1986representations}
\begin{equation}
\label{eq:orderexp}
|h_i|^{2}=\sum\limits_{k=1}^{i}y_k,
\end{equation}
\noindent where the $y_k$'s are independent exponential random variables with rate parameter $\frac{1}{M+1-k}$. Thus, substituting (\ref{eq:orderexp}) in  (\ref{eq:cond_step1}) and isolating $y_1$ yields 
\begin{equation}
\label{eq:E_i}
E_{i|s}^c=\left\{y_1>\frac{\frac{\phi_i}{\rho_s}+\phi_i\sum\limits_{q=2}^{i-1}a_q\sum\limits_{p=2}^{q}y_p-a_i\sum\limits_{l=2}^{i}y_l}{a_i-\phi_i\sum\limits_{q=1}^{i-1}a_q} \right\}.
\end{equation}
\noindent Therefore, $\textup{Pr}\left(E_{i|s}^c\right)$ conditioned on $y_2,...,y_i$ is formulated as
\small
\begin{equation}
\label{eq:intE}
   \textup{Pr}\left( E_{i|s}^c|y_2,...,y_i\right)= 1-F_{y_1}\left(\frac{\frac{\phi_i}{\rho_s}+\phi_i\sum\limits_{q=2}^{i-1}a_q\sum\limits_{p=2}^{q}y_p-a_i\sum\limits_{l=2}^{i}y_l}{a_i-\phi_i\sum\limits_{q=1}^{i-1}a_q}\right ),
\end{equation}
\normalsize
\small
\begin{figure*}[ht]
\begin{equation}
\label{eq:final_E}
    \textup{Pr}\left( E_{i|s}^c\right)=\left\{\begin{matrix}
1-\int_{0}^{\xi_i}...\int_{0}^{\xi_2}\left(1-e^{-M\frac{\frac{\phi_j}{\rho_s}+\phi_j\sum\limits_{q=2}^{i-1}a_q\sum\limits_{p=2}^{q}y_p-a_i\sum\limits_{l=2}^{i}y_l}{a_i-\phi_j\sum\limits_{q=1}^{i-1}a_q}}\right) \left(M-1\right)e^{-\left(M-1\right)y_2}...&\\... \times \left(M+1-i\right)e^{-\left(M+1-i\right)y_i} dy_2 ... dy_i, & \textup{if}\,a_i>\phi_j\sum\limits_{q=1}^{i-1}a_q\\ 
 1-\int_{0}^{\xi_i}...\int_{0}^{\xi_2}\left(M-1\right)e^{-\left(M-1\right)y_2}... \left(M+1-i\right)e^{-\left(M+1-i\right)y_i} dy_2 ... dy_i, & \textup{if}\, a_i=\phi_j\sum\limits_{q=1}^{i-1}a_q\\ 
1- \int_{\xi_i}^{\infty}...\int_{\xi_2}^{\infty}\left(1-e^{-M\frac{\frac{\phi_j}{\rho_s}+\phi_j\sum\limits_{q=2}^{i-1}a_q\sum\limits_{p=2}^{q}y_p-a_i\sum\limits_{l=2}^{i}y_l}{a_i-\phi_j\sum\limits_{q=1}^{i-1}a_q}}\right )\left(M-1\right)e^{-\left(M-1\right)y_2}...&\\ ...\times \left(M+1-i\right)e^{-\left(M+1-i\right)y_i} dy_2 ...dy_i, &\textup{if}\, a_i<\phi_j\sum\limits_{q=1}^{i-1}a_q
\end{matrix}\right.
\end{equation}
\hrulefill
\vspace*{-10pt}
\end{figure*}
\normalsize
\noindent where $F_{Y_1}\left(y \right)$ denotes the cumulative distribution function (CDF) of the exponential random variable $y_1$. Integrating over $y_2,...,y_i$'s respective distributions, we obtain (\ref{eq:final_E}) at the top of the next page, where the bounds of the integrals are given by
\begin{equation}
 \xi_{k}=\left\{\begin{matrix}
 \frac{\frac{\phi_i}{\rho_s}+\phi_i\sum\limits_{q=k+1}^{i-1}a_q\sum\limits_{p=k+1}^{q}y_p-a_i\sum\limits_{l=k+1}^{i}y_l}{a_i-\phi_i\sum\limits_{q=k}^{i-1}a_q},& 2\leq k<i \\ 
 \frac{\phi_i}{\rho_s a_i},& k=i
\end{matrix}\right.
\end{equation}
\noindent It is worth mentioning that the derived OP of an $M$ SUs uplink NOMA scenario with i.i.d. ordered channels, due to it's complexity \cite{7390209}, has not been reported in the open technical literature, even for the Gaussian noise case. However, given the identity in (\ref{eq:orderexp}), we obtained the OP for $M$ SUs in terms of simple integrals of exponential functions that can be evaluated given $M$. Furthermore, without loss of generality, in what follows we provide the closed form expression for a $3$ SUs NOMA system under impulsive noise. For $M=3$, substituting $j=1$, $2$ and $3$ in (\ref{eq:intE}), we get
\begin{equation}
\begin{split}
\label{eq:e1}
    \textup{Pr}\left( E_{1|s}^c\right)&=1-F_{y_1}\left(\frac{\phi_1}{\rho_s a_1}\right)\\&=e ^{-\frac{3\phi_1}{\rho_s a_1} },
    \end{split}
\end{equation}
\begin{equation}
\begin{split}
\label{eq:E_2}
    \textup{Pr}\left( E_{2|s}^c|y_2\right)
    &=1-F_{y_1}\left(\frac{\frac{\phi_2}{\rho_s}-a_2y_2}{a_2-\phi_2a_1}\right),
\end{split}
\end{equation}
\noindent and 
\small
\begin{equation}
\begin{split}
\label{eq:E_3}
    \textup{Pr}\left( E_{3|s}^c|y_2,y_3\right)
    =1-F_{y_1}\left(\frac{\frac{\phi_3}{\rho_s}+\phi_3a_2y_2-a_3\left(y_2+y_3\right)}{a_3-\phi_i\left(a_1+a_2\right)}\right).
    \end{split}
\end{equation}
\normalsize
\noindent Consequently, for $a_2>\phi_2a_1$, the unconditional $\textup{Pr}\left( E_{2|s}^c\right)$ is formulated as 
\small
\begin{equation}
\label{eq:E_2int}
\textup{Pr}\left( E_{2|s}^c\right)=1-
\int\limits_{0}^{\frac{\phi_2}{\rho_s a_2}} \left(1-e^{-3 \frac{\frac{\phi_2}{\rho_s}-a_2y_2}{a_2-\phi_2a_1}}\right )2e^{-2y_1 }dy_1.
\end{equation}
\normalsize
\noindent  Solving the integral yields
\small
\begin{equation}
\label{eq:e2}
    \textup{Pr}\left( E_{2|s}^c\right)=\exp\left(\frac{-2\phi_2}{\rho_sa_2}\right)+\frac{\exp\left(\frac{-3\phi_2}{\rho_s\left(a_2-\phi_2a_1 \right )} \right )-\exp\left(\frac{-2\phi_2}{\rho_sa_2} \right )}{1-\frac{3a_2}{2\left(a_2-\phi_2a_1\right)}}.
\end{equation}
\normalsize
Similarly, for $a_3>\phi_3\left(a_1+a_2 \right )$, $\textup{Pr}\left( E_{3|s}^c\right)$ is formulated as
\small
\begin{equation}
\begin{split}
\label{eq_E_3int}
\textup{Pr}\left( E_{3|s}^c\right)=1-
\int\limits_{0}^{c_3}\int\limits_{0}^{c_1}&\left(1-e^{-3\left(\frac{\frac{\phi_3}{\rho_s}+\phi_3a_2y_2-a_3\left(y2+y3 \right )}{c_2} \right )}\right)\\&\times2 \,e^{-2y_2 }e^{-y_1 }dy_2\,dy_1, 
\end{split}
\end{equation}
\normalsize
\noindent where $c_1=a_3- \phi_3 \,a_2$, $c_2=a_3-\phi_3\left(a_1+a_2\right)$, and $c_3=\frac{\phi_3}{\rho_s a_3}$. Solving the integrals yields
\small
\begin{equation}
\label{eq:e3}
\begin{split}
    &\textup{Pr}\left( E_{3|s}^c\right)=e^{-c_3 }+e^{\frac{-2\phi_3}{\rho_s c_1} }\frac{1-e^{-c_3\left(1-\frac{2a_3}{c_1} \right)  }}{1-\frac{2a_3}{c_1}}+\frac{2e^{\frac{-3\phi_3}{\rho_s c_2} }}{2-\frac{3c_1}{c_2}}
\\ & \times \left(\frac{1-e^{-c_3\left(1-\frac{3a_3}{c_2} \right ) }}{1-\frac{3a_3}{c_2} }-\frac{1-e^{-c_3\left(1-\frac{3a_3}{c_2}-\frac{a_3\left(2-\frac{3c_1}{c_2} \right )}{c_1}\right ) }}{e^{\frac{\phi_3\left(2-\frac{3c_1}{c_2} \right )}{\rho_s c_1}}\left(1-\frac{3a_3}{c_2}-\frac{a_3\left(2-\frac{3c_1}{c_2} \right )}{c_1}\right)}  \right ),
 \end{split}
\end{equation}
\normalsize
\noindent Therefore, substituting \eqref{eq:e1}, \eqref{eq:e2}, and \eqref{eq:e3} in \eqref{eq:POUT} yields the OP of the different SUs.
\section{Asymptotic Analysis}
\label{sec:asymptotic}
Capitalizing on the derived outage probability expressions, in this section we proceed to analyze the effect of impulsive noise on NOMA systems by quantifying its effect on the asymptotic diversity order of such systems. The asymptotic diversity order characterizes the magnitude of the slope of the OP as a function of the average SNR in a log-log scale \cite{1603372}. Thus, the asymptotic diversity order of uplink NOMA under impulsive noise is evaluated as \cite{8327577} 
\begin{equation}
\label{eq:diversity}
d_a=\lim\limits_{\rho_w\rightarrow \infty}\frac{-\log P_{{\rm out}}}{\log\rho_w}
\end{equation}
\noindent Since $\exp\left({-x} \right) \underrel{x\rightarrow  0}{\approx} 1-x$, substituting \eqref{eq:POUT} in \eqref{eq:diversity}, yields
\begin{equation}
\begin{split}
\label{eq:div_P1}
d_a&(P_{(out,1)})\approx\lim\limits_{\rho_w\rightarrow \infty}\frac{\log \left( \frac{\rho_w a_1 }{3 \phi_1 \left(1-p-\frac{p}{\left(\Gamma+1\right)}\right)}\right )}{\log \left( \rho_w\right)}\\
&\approx\lim\limits_{\rho_w\rightarrow \infty}\frac{\log \left(\rho_w  \right)+\log \left(a_1 \right)-\log \left(3 \phi_1 \left(1-p-\frac{p}{\left(\Gamma+1\right)}\right) \right)}{\log \left( \rho_w\right)},
\end{split}
\end{equation}
\begin{equation}
\begin{split}
\label{eq:div_P2}
d_a(P_{(out,2)})&\approx\lim\limits_{\rho_w\rightarrow \infty}\frac{\log \left( \frac{\rho_w^2 a_2 \left(a_2-\phi_2 a_1 \right )}{6 \phi_2^2 \left(\left(1-p\right)-\frac{p}{\left(\Gamma+1\right)^2}\right)}\right )}{\log \left( \rho_w\right)}\\ &\approx\lim\limits_{\rho_w\rightarrow \infty}\frac{2\log \left(\rho_w  \right)+\log \left(a_2 \left(a_2-\phi_2 a_1 \right )\right)}{\log \left( \rho_w\right)}\\&\,\,\,\,\,\,\,\,\,\,\,\,\,\,\,\,\,\,\;\;\;-\frac{\log \left(6 \phi_2^2 \left(1-p-\frac{p}{\left(\Gamma+1\right)^2}\right) \right)}{\log \left( \rho_w\right)},
\end{split}
\end{equation}
\begin{equation}
\begin{split}
\label{eq:div_P3}
d_a(P_{(out,3)})&\approx\lim\limits_{\rho_w\rightarrow \infty}\frac{\log \left( \frac{\rho_w^3 a_3 c_1 c_2}{6 \Phi_3^3 \left(\left(1-p\right)-\frac{p}{\left(\Gamma+1\right)^3}\right)}\right )}{\log \left( \rho_w\right)}\\
&\approx\lim\limits_{\rho_w\rightarrow \infty}\frac{3\log \left(\rho_w  \right)+\log \left(a_3 c_1 c_2\right)}{\log \left( \rho_w\right)}\\&\,\,\,\,\,\,\,\,\,\,\,\,\,\,\,\,\,\,\;\;\;-\frac{\log \left(6 \phi_2^2 \left(1-p-\frac{p}{\left(\Gamma+1\right)^3}\right) \right)}{\log \left( \rho_w\right)},
\end{split}
\end{equation}
\noindent and consequently
\begin{equation}
    d_a(P_{(out,j)})\approx j.
\end{equation}
\noindent This implies that the impulsive noise does not affect the asymptotic diversity order of the implied NOMA users. However, it does delay its appearance due to the $\left(1-p-\frac{p}{\left(\Gamma+1\right)^j}\right)$ that affects the last log in \eqref{eq:div_P1}-\eqref{eq:div_P3}. This term, which comes from the impulsive noise and is included in $[0 , 1]$, decreases the value of the last log, which increases the $\rho_w$ required in order to reach the asymptotic regime. Hence, this analysis suggests that such noise will shift the appearance of the asymptotic regime and consequently degrade the performance of the associated NOMA users.  
\section {Numerical and Simulation Results}
Considering the aforementioned NOMA approach, this section investigates the effect of impulsive noise on the performance of uplink NOMA-based communication systems. To this end, extensive Monte Carlo simulations have been executed in order to validate the derived analytical results and investigate the OP performance of NOMA systems subject to impulsive noise. Without loss of generality, it is noted that the number of SUs considered is $M=3$ and that the target data rate ${R^t_1}={R^t_2}={R^t_3}=0.5 \rm{bits/s/Hz}$. Here, the numerical results are represented by solid lines, whereas markers are used to illustrate the corresponding computer simulation results. In this regard, it is observed that the derived analytical results perfectly match the corresponding computer simulations.
\begin{figure}[ht]
\centering
\includegraphics[width=1\linewidth]{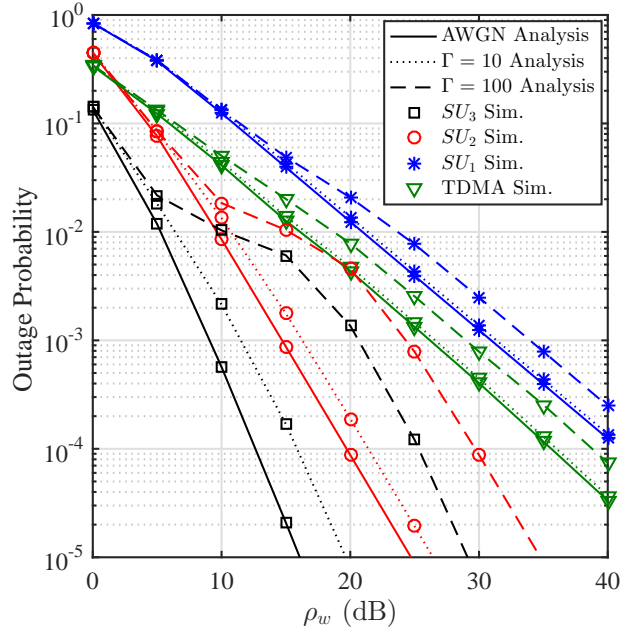}
\caption{Analytical and Simulation results of the OP versus $\rho_w$ for a $3$ SUs NOMA and TDMA system subject to impulsive noise with $p=0.01$.}
\label{fig:Fig1}
\end{figure}
\begin{figure}[ht]
\centering
\includegraphics[width=1\linewidth]{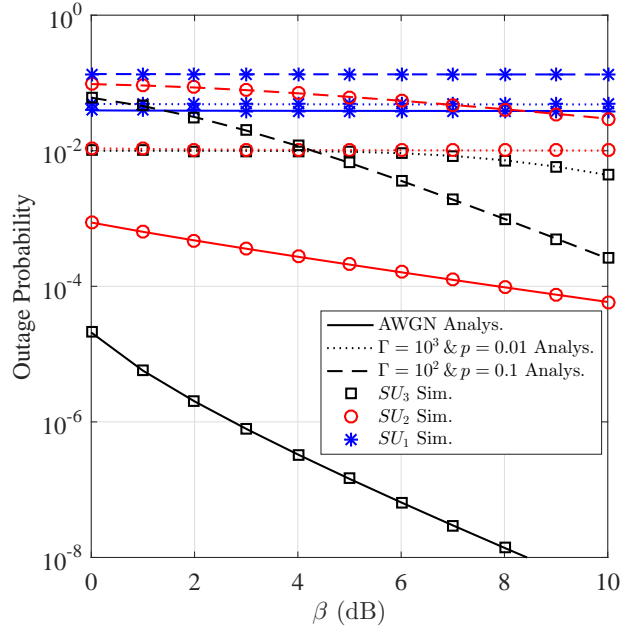}
\caption{Analytical and Simulation results of the OP versus back-off coefficient $\beta$ for a $3$ SUs NOMA system subject to impulsive noise with $\rho_w=15$ dB.}
\label{fig:Fig2}
\end{figure}

Fig. \ref{fig:Fig1} shows the effect of impulsive noise on the OP performance of a $3$ SUs NOMA-based system compared to a time division multiple access (TDMA)-based system. Without loss of generality, here, the transmit power of all the users is set to unity ($a_1=a_2=a_3=1$).  First, it is observed that this type of noise causes significant performance losses to the NOMA SUs and that the level of performance degradation induced depends on the order of the SU. This is consistent with the results obtained in Section \ref{sec:asymptotic} where it was shown that the impulsive noise does not affect the diversity order of the SUs, but rather shifts the appearance of the asymptotic regime and this shift depends on the order of the considered user $j$. Specifically, for $\Gamma=100$, we can observe a loss of more than $10$ dB for $SU_3$ and $SU_2$ against around $3$ dB for $SU_1$. This can be explained by the fact that higher orders of SUs, i.e., less SIC operations performed, benefit from better channel conditions, but they also experience interference from the lower order users; therefore, they are quite sensitive to any additional noise/interference. In addition, it is observed that, even for very low impulsive noise power and occurrence rate, which does not affect the performance of the TDMA SUs, NOMA SUs are not spared. Specifically, the performance of the higher order SUs, is affected and this degradation is expected to increase as the number of SUs increases.

In Fig. \ref{fig:Fig2}, we set a power back-off coefficient $\beta$, which implies that the transmit signal power of the $i^{\rm th}$ user is $\beta$ dB stronger than that of user $i-1$. Here, we consider a relatively highly impulsive ($p=0.01$ and $\Gamma=10^3$) as well as a lower impulsiveness ($p=0.1$ and $\Gamma=10^2$) scenarios where, without loss of generality, we set $\rho_w=15$ dB and $a_1=1$. This implies that the horizontal axis represents the SNR of the weakest SU. Interestingly, it is observed that, for the highly impulsive scenario, the introduction of a back-off coefficient practically doesn't affect the performance of the system, where for $\beta \leq 6$ there is no improvement in the OP of the different users and thus, the power increase associated to the use of a back-off coefficient is inefficient. This is because the impulsive noise outweighs the interference from the NOMA users in this case.
\section{Conclusion}
In this paper, we investigated the performance of NOMA systems subject to impulsive noise. Specifically, we considered the case of uplink transmission where $M$ SUs simultaneously communicate with an AP under the effect of both fading and impulsive noise. We considered the widely used Bernoulli-Gaussian model to characterize the non Gaussian impulses. Our findings suggest that, even for very low impulsive noise power and occurrence rate, NOMA users, and especially the higher order ones, are quite sensitive to impulsive noise. Precisely, it has been demonstrated that the inherent nature of NOMA renders its users particularly sensitive to impulsive interference. This is due the presence of interference from the non-orthogonal multiplexing of the users in the power domain which increases the vulnerability of the system. In addition, this nonorthogonal multiplexing is also expected to affect the performance of conventional mitigation techniques, designed for OMA schemes. Therefore, the analysis, mitigation, and optimization of NOMA-based systems under the effects of impulsive noise should be addressed prior to the deployment of NOMA in impulsive noise environments.
\balance
\bibliographystyle{IEEEtran}
\bibliography{references}
\end{document}